 \documentclass[onecolumn,12pt,draft]{IEEEtran} 

\usepackage{epsf}

\begin{document}

\title{ 
Analysis of ``SIR'' (``Signal''-to-``Interference''-Ratio) in Discrete-Time 
Autonomous Linear Networks with Symmetric Weight Matrices
} 

\author{Zekeriya Uykan \thanks{Z. Uykan is with Helsinki University of Technology, Control
Engineering Laboratory, FI-02015 HUT, Finland, and Nokia Siemens Networks, Espoo, Finland. 
E-mail: zekeriya.uykan@hut.fi, zuykan@seas.harvard.edu.
The author has been a visiting scientist at Harvard University Broadband Comm Lab., Cambridge, 02138\
, MA,
since September 2008
and this work has been performed during his stay at Harvard University.
} 
}


\maketitle
\begin{abstract}

It's well-known that in a traditional discrete-time autonomous linear systems, 
the eigenvalues of the weigth (system) matrix solely determine the stability of the system.  
If the spectral radius of the system matrix is larger than 1, then the system is unstable. In this 
paper, we examine the linear systems 
with symmetric weight matrix whose spectral radius is larger than 1. 

The author introduced a dynamic-system-version of "Signal-to-Interference Ratio (SIR)" 
in nonlinear networks in \cite{Uykan08a} and \cite{Uykan08b} and in 
continuous-time linear networks in \cite{Uykan09a}. 
Using the same "SIR" concept, 
we, in this paper, analyse the "SIR" of the states in the following two 
$N$-dimensional discrete-time autonomous linear systems: 
1) The system 
${\mathbf x}(k+1) = \big( {\bf I} + \alpha ( -r {\bf I} + {\bf W} ) \big) {\mathbf x}(k)$ 
which is obtained by discretizing the autonomous continuous-time 
linear system in \cite{Uykan09a} using Euler method; where 
${\bf I}$ is the identity matrix, $r$ is a positive real number, and 
$\alpha >0$ is the step size. 
2) A more general autonomous linear system descibed by ${\mathbf x}(k+1) = -\rho {\mathbf I + W} {\mathbf x}(k)$, 
where ${\mathbf W}$ is any real symmetric matrix whose diagonal elements are zero, and 
${\bf I}$ denotes the identity matrix and $\rho$ is a positive real number. 
Our analysis shows that: 1) The "SIR" of any state 
converges to a constant value, called "Ultimate SIR", in a finite time 
in the above-mentioned discrete-time linear systems. 
2) The "Ultimate SIR" in the first system above is equal to $\frac{\rho}{\lambda_{max}}$ where 
$\lambda_{max}$ is the maximum (positive) eigenvalue of the matrix ${\bf W}$. 
These results are in line with those of \cite{Uykan09a} where corresponding continuous-time linear system is examined. 
3)  The "Ultimate SIR" in the second system above is equal to $\frac{\rho}{\lambda_{m}}$ where 
$\lambda_{m}$ is the eigenvalue of ${\bf W}$ which satisfy 
$| \lambda_{m} - \rho | = \max  \{ | \lambda_{i} - \rho | \}_{i=1}^{N}$ if  
$\rho$ is accordingly determined from the interval $0 < \rho < 1$.

In the later part of the paper, we use the introduced "Ultimate SIR" to stabilize the (originally unstable) networks. 
It's shown that the proposed Discrete-Time 
"Stabilized"-Autonomous-Linear-Networks-with-Ultimate-SIR" exhibit features which are 
generally attributed to Discrete-Time Hopfield Networks. 
Taking the sign of the converged states, the proposed networks are applied to binary associative memory design. 
Computer simulations show the effectiveness of the 
proposed networks as compared to traditional discrete Hopfield Networks.

\end{abstract}


\section{Introduction \label{Section:INTRO}}

\PARstart{S}{ignal}-to-Interference Ratio (SIR) is an important entity in commucations engineering 
which indicates the quality of a link between a transmitter and a receiver in a multi transmitter-receiver 
environment (see e.g. \cite{Rappaport96}, among many others).  
For example, let $N$ represent the number of transmitters and receivers using the same channel. Then the received 
SIR at receiver $i$ is given by (see e.g. \cite{Rappaport96}) 

\begin{equation} \label{eq:cir}
SIR_i(k) = \gamma_i(k) = \frac{ g_{ii} p_i(k)}{ \nu_i + \sum_{j = 1, j \neq i}^{N} g_{ij} p_j(k) },  \quad i=1, \dots, N
\end{equation}

where  $p_i(k)$ is the transmission power of transmitter $i$ at time step $k$, $g_{ij}$ is
the link gain from transmitter $j$ to receiver $i$ (e.g. in case of wireless communications, 
$g_{ij}$ involves path loss, shadowing, etc) and $\nu_i$ represents the
receiver noise at receiver $i$. 
Typically, in wireless communication systems like cellular radio systems, every transmitter tries to optimize its 
power $p_i(k)$ such that the received SIR(k) (i.e., $\gamma_i(k)$) in eq.(\ref{eq:cir}) is kept at a 
target SIR value, $\gamma_i^{tgt}$. In an interference dominant scenario, the receiver background noise $\nu_i$ 
can be ignored and then 

\begin{equation} \label{eq:cir_Intdominant}
\gamma_i(k) = \frac{ g_{ii} p_i(k)}{ \sum_{j = 1, j \neq i}^{N} g_{ij} p_j(k) },  \quad i=1, \dots, N
\end{equation}

The author defines the following dynamic-system-version of ``Signal-to-Interference-Ratio (SIR)'', 
denoted by $\theta_i(k)$, by rewriting the eq.(\ref{eq:cir}) with neural network terminology 
in \cite{Uykan08a} and \cite{Uykan08b}: 

\begin{equation} \label{eq:cirA}
{\theta_i}(k) = \frac{ a_{ii} x_i(k)}{ \nu_i + \sum_{j = 1, j \neq i}^{N} w_{ij} x_j(k) },  
	\quad i=1, \dots, N
\end{equation}

where  
$\theta_i(k)$ is the defined ficticious ``SIR'' at time step $k$, 
$x_i(k)$ is the state of the $i$'th neuron, 
$a_{ii}$ is the feedback coefficient from its state to its input layer,  
$w_{ij}$ is the weight from the output of the $j$'th neuron to the input of the 
$j$'th neuron.  For the sake of brevity, in this paper, 
we assume the "interference dominant" case, 
i.e. $\nu_i$ is negligible. 

A traditional discrete-time autonomous linear network is given by  

\begin{equation} \label{eq:Diff_Linear_trad} 
{\mathbf x}(k+1) = {\mathbf M} {\mathbf x}(k), \quad \quad \quad {\mathbf x}(k) \in R^{N \times 1}, \quad 
{\mathbf M} \in R^{N \times N},
\end{equation} 

where ${\mathbf x}(k)$ shows the state vector 
at time $t$ and square matrix ${\mathbf M}$ is called 
system matrix or weight matrix. 

It's well-known that in the system of eq.(\ref{eq:Diff_Linear_trad}), 
the eigenvalues of the weight (system) matrix solely determine the stability of system.  
If the spectral radius of the matrix is larger than 1, then the system is unstable. In this 
paper, we examine the linear systems 
with system matrices whose spectral radius is larger than 1.

Using the same "SIR" concept in eq.(\ref{eq:cirA}), 
we, in this paper, analyse the "SIR" of the states in the following two 
discrete-time autonomous linear systems: 

\begin{enumerate}

\item  
The following linear system which is obtained by discretizing the continuous-time 
linear system 
$\dot{ {\mathbf x}} =  \Big( -r {\bf I} + {\mathbf W} \Big) {\bf x}$
in \cite{Uykan09a} using Euler method:

\begin{equation} \label{eq:DiscretizedLS}
{\mathbf x}(k+1) = \Big( {\bf I} + \alpha ( -r {\bf I} + {\bf W} ) \Big) {\mathbf x}(k) 
\end{equation}

where ${\bf I}$ denotes the identity matrix, $(-r {\bf I} + {\bf W})$ is the real symmetric 
system matrix with zero-diagonal ${\bf W}$, and $\alpha$ is the step size.

\item 
A more general autonomous linear system descibed by 

\begin{equation} \label{eq:DiscLS_general}
{\mathbf x}(k+1) = ( -r {\mathbf I + W} ) {\mathbf x}(k)
\end{equation}

where ${\mathbf W}$ is any real symmetric matrix whose diagonal elements are zero, and 
${\bf I}$ denotes the identity matrix and $r$ is a positive real number. 

\end{enumerate}

Our analysis shows that: 1) The ``SIR'' of any state
converges to a constant value, called ``Ultimate SIR'', in a finite time
in the above-mentioned discrete-time linear systems.
2) The ``Ultimate SIR'' in the first system above is equal to $\frac{\rho}{\lambda_{max}}$ where
$\lambda_{max}$ is the maximum (positive) eigenvalue of the matrix ${\bf W}$.
These results are in line with those of \cite{Uykan09a} where corresponding continuous-time linear system is examined.
3)  The ``Ultimate SIR'' in the second system above is equal to $\frac{\rho}{\lambda_{m}}$ where
$\lambda_{m}$ is the eigenvalue of ${\bf W}$ which satisfy
$| \lambda_{m} - \rho | = \max  \{ | \lambda_{i} - \rho | \}_{i=1}^{N}$ if
$\rho$ is accordingly determined from the interval $0 < \rho < 1$.

In the later part of the paper, we use the introduced "Ultimate SIR" to stabilize the (originally unstable) network. 
It's shown that the proposed Discrete-Time 
"Stabilized"-Autonomous-Linear-Networks-with-Ultimate-SIR" 
exhibit features which are 
generally attributed to Discrete-Time Hopfield Networks. 
Taking the sign of the converged states, the proposed networks are applied to binary associative memory design. 

The paper is organized as follows:  The ultimate "SIR" is analysed for the 
autonomous linear discrete-time systems with symmetric weight matrices in section 
\ref{Section:USIR}.  Section \ref{Section:proposedNet} presents the stabilized networks by their Ultimate "SIR" 
to be used as a binary associative memory system. 
Simulation results are presented in section \ref{Section:SimuResults}, which is followed by the conclusions in 
Section \ref{Section:CONCLUSIONS}.

\vspace{0.2cm}

\section{Analysis of ``SIR'' in Discrete-Time Autonomous Linear Networks with Symmetric Weight Matrices}
\label{Section:USIR}

In this section, we analyse the ``SIR'' of the states in the following two
discrete-time autonomous linear systems:
1) The discrete-time autonomous system which is obtained by discretizing the continuous-time
linear system in \cite{Uykan09a} using Euler method; and
2) A more general autonomous linear system descibed by ${\mathbf x}(k+1) = (-\rho {\bf I} + {\bf W}) {\mathbf x}(k)$,
where ${\mathbf W}$ is any real symmetric matrix whose diagonal elements are zero, and
${\bf I}$ denotes the identity matrix and $r$ is a positive real number.

\subsection{ Discretized Autonomous Linear Systems with symmetric matrix case \label{Subsection:Mx} }

The author examines the ``SIR'' in the following continuous-time linear system in \cite{Uykan09a}:

\begin{equation} \label{eq:Diff_Linear_cont}
\dot{ {\mathbf x}} =  \Big( -r {\mathbf I} + {\mathbf W} \Big) {\mathbf x}
\end{equation}

where $\dot{ {\mathbf x}}$ shows the derivative of ${\mathbf x}$ with respect to time, i.e.,
$\dot{ {\mathbf x}} = \frac{d{\mathbf x}}{dt}$.
In this subsection, we analyse the following discrete-time autonomous linear system which 
is obtained by discretizing the continuous-time system of eq.(\ref{eq:Diff_Linear_cont})
by using well-known Euler method: 

\begin{equation} \label{eq:Diff_Linear}
{\mathbf x}(k+1) = \big( {\bf I} + \alpha ( -r {\bf I} + {\bf W} ) \big) {\mathbf x}(k)
\end{equation}

where ${\bf I}$ is the identity matrix, $r$ is a positive real number, 
$(-r {\bf I} + {\bf W})$ is the system matrix, 
${\mathbf x}(k)$ shows the state vector at step $k$, and 
$\alpha >0$ is the step size and   

\begin{equation} \label{eq:matA_W_b}
r {\bf I} =
\left[
\begin{array}{c c c c}
r   &   0   & \ldots  &  0 \\
0     &   r & \ldots  &  0 \\
\vdots &      & \ddots  &  0 \\
0     &   0   & \ldots  &  r
\end{array}
\right]_{N \times N}, 
\quad \quad
{\mathbf W} =
\left[
\begin{array}{c c c c}
0  &   w_{12}   & \ldots  &  w_{1N} \\
w_{21}     &   0 & \ldots  &  w_{2N} \\
\vdots &     & \ddots  &  \vdots \\
w_{N1}    &   w_{N2}   & \ldots  &  0
\end{array}
\right]_{N \times N}
\end{equation}

In this paper, we examine only the linear systems with symmetrix weight matrices, i.e., 
$w_{ij}=w_{ji}, \quad i,j=1,2, \dots, N$. 
It's well known that desigining the weight matrix ${\mathbf W}$ as a symmetric one yields that all 
eigenvalues are real (see e.g. \cite{Bretscher05}),
which we assume throughout the paper due to the simplicity and brevity of its analysis. 

The reason of the notation in (\ref{eq:matA_W_b}) is because we prefer to have 
the same notation as in \cite{Uykan08a}. 


\vspace{0.2cm}
\emph{Proposition 1:} 
\vspace{0.2cm}

In the autonomous discrete-time linear network of eq.(\ref{eq:Diff_Linear}), 
let's assume that the spectral radius of the system  matrix 
$\big( {\bf I} + \alpha ( -r {\bf I} + {\bf W} ) \big)$ is larger than 1. 
(This assumption is equal to the assumption that 
${\bf W}$ has positive eigenvalue(s) and 
$r>0$ is chosen such that $\lambda_{max} > r$, where 
$\lambda_{max}$ is the maximum (positive) eigenvalue of ${\bf W}$).  
If $\alpha$ is chosen such that $0 < \alpha r < 1$, then 
the defined "SIR" ${\theta_i}(k)$ in eq.(\ref{eq:cirA}) for any state $i$  
converges to the following constant within a finite step number 
for any initial vector ${\bf x}(0)$ which is 
not completely perpendicular to the eigenvector corresponding to the largest eigenvalue of ${\bf W}$. 
\footnote{
It's easy to check in advance if the initial vector ${\bf x}(0)$ is 
completely perpendicular to the eigenvector of the maximum (positive) eigenvalue of ${\bf W}$ or not. 
If this is the case, then this can easily be overcome 
by introducing a small random variable to ${\bf x}(0)$ so that it's not completely perpendicular to the 
mentioned eigenvector.
} 

\begin{equation} \label{eq:theta_const}
\theta_i(k \geq k_T) =  \frac{r}{ \lambda_{max} }, \quad \quad i=1,2, \dots, N
\end{equation} 

where $\lambda_{max}$ is the maximum (positive) eigenvalue of the weight matrix ${\bf W}$ 
and $k_T$ shows a finite time constant.

\begin{proof}

From eq. (\ref{eq:Diff_Linear}), it's obtained  

\begin{equation} \label{eq:solution}
{\mathbf x}(k) = \Big( {\bf I} + \alpha ( -r {\bf I} + {\bf W} ) \Big)^{k} {\mathbf x}(0) 
\end{equation}

where ${\bf x}(0)$ shows the initial state vector at step zero. Let us first examine the powers of the matrix 
$\Big( {\bf I} + \alpha ( -r {\bf I} + {\bf W} ) \Big)$ in (\ref{eq:solution}) 
in terms of matrix $r{\bf I}$ and the eigenvectors of matrix ${\bf W}$: 

It's well known that any symmetric real square matrix can be decomposed into 

\begin{equation} \label{eq:symW}
{\bf W} = \sum_{i=1}^{N} \lambda_i {\bf v}_i {\bf v}_i^T = \sum_{i=1}^{N} \lambda_i {\bf V}_i 
\end{equation}

where $\{ \lambda_i \}_{i=1}^{N}$ and $\{ {\bf v}_i \}_{i=1}^{N}$ show the (real) 
eigenvalues and the corresponding eigenvectors and 
the eigenvectors $\{ {\bf v}_i \}_{i=1}^{N}$ are orthonormal (see e.g. \cite{Bretscher05}), i.e., 

\begin{equation} \label{eq:v_ortnor}
{\bf v}_i {\bf v}_j = 
\left\{ 
\begin{array}{ll}
1 & \textrm{if} \quad i=j, \quad \quad \textrm{where} \quad i,j=1,2,\dots,N \\
0 & \textrm{if} \quad i \neq j, 
\end{array}
\right.
\end{equation}

Let's define the outer-product matrices of the eigenvectors  $\{ \lambda_i \}_{i=1}^{N}$ as 
${\bf V}_j = {\bf v}_i {\bf v}_i^T, \quad i=1,2, \dots, N$, which, from eq.(\ref{eq:v_ortnor}), is equal to 

\begin{equation} \label{eq:V_ortnorjMat}
{\bf V}_j = 
\left\{ 
\begin{array}{ll}
{\bf I} & \textrm{if} \quad i=j, \quad \quad \textrm{where} \quad i,j = 1,2,\dots,N \\
{\bf 0} & \textrm{if} \quad i \neq j, 
\end{array}
\right.
\end{equation}

where ${\bf I}$ is the identity matrix. Defining matrix ${\bf M}$, 

\begin{equation} \label{eq:M}
{\bf M} = {\bf I} + \alpha ( -r {\bf I} + {\bf W} )
\end{equation}

which is obtained as 

\begin{equation} \label{eq:W1}
{\bf M} = (1 - \alpha r) {\bf I} + \sum_{i=1}^{N} \beta_i(1) {\bf V}_i
\end{equation}

where $r > 0$, $\alpha > 0$, and where $\beta_i(1)$ is equal to 

\begin{equation} \label{eq:beta1}
\beta_i(1) = \alpha \lambda_i
\end{equation}

The matrix ${\bf M}^2$ can be written as 

\begin{equation} \label{eq:W2}
{\bf M}^2 = (1 - \alpha r)^2 {\bf I} + \sum_{i=1}^{N} \beta_i(2) {\bf V}_i
\end{equation}

where $\beta_i(2)$ is equal to 

\begin{equation} \label{eq:beta2} 
\beta_i(2) = \alpha (1-\alpha r) \lambda_i + (1-\alpha r + \alpha \lambda_i) \beta_i(1) 
\end{equation} 

Similarly, the matrix ${\bf M}^3$ can be written as 

\begin{equation} \label{eq:W3}
{\bf M}^3 = (1 - \alpha r)^3 {\bf I} + \sum_{i=1}^{N} \beta_i(3) {\bf V}_i
\end{equation}

where $\beta_i(3)$ is equal to 

\begin{equation} \label{eq:beta3}
\beta_i(3) = \alpha (1-\alpha r)^2\lambda_i + (1-\alpha r + \alpha \lambda_i) \beta_i(2)
\end{equation}

So, ${\bf M}^4$ can be written as 

\begin{equation} \label{eq:W4}
{\bf M}^4 = (1 - \alpha r)^4 {\bf I} + \sum_{i=1}^{N} \beta_i(4) {\bf V}_i
\end{equation}

where $\beta_i(4)$ is equal to 

\begin{equation} \label{eq:beta4}
\beta_i(4) = \alpha (1-\alpha r)^3\lambda_i + (1-\alpha r + \alpha \lambda_i) \beta_i(3)
\end{equation}

So, at step $k$, the matrix $({\bf M})^k$ is obtained as 

\begin{equation} \label{eq:W_k}
{\bf M}^k = (1 - \alpha r)^k {\bf I} + \sum_{i=1}^{N} \beta_i(k) {\bf V}_i
\end{equation}

where $\beta_i(k)$ is equal to 

\begin{equation} \label{eq:beta_i_k}
\beta_i(k) = \alpha (1-\alpha r)^{k-1} \lambda_i + ( 1 + \alpha ( \lambda_i - r) ) \beta_i(k-1)
\end{equation}

Using eq.(\ref{eq:beta1}) and (\ref{eq:beta_i_k}), the $\beta_i(k)$ is obtained as 

\begin{eqnarray}  
\beta_i(1) &  = & \alpha \lambda_i   \label{eq:betasum1} \\
\beta_i(2) &  = & \alpha \lambda_i \Big( (1-\alpha r) + \big( 1 + \alpha ( \lambda_i - r) \big) \Big) 
		  \label{eq:betasum2} \\
\beta_i(3) &  = & \alpha \lambda_i \Big( (1-\alpha r)^2 + (1-\alpha r) \big( 1 + \alpha ( \lambda_i - r) \big) + 
\big( 1 + \alpha ( \lambda_i - r) \big)^2 \Big)  \label{eq:betasum3} \\ 
    	   &  \vdots  &    \label{eq:betasumdots}    \\
\beta_i(k)   & = & \alpha \lambda_i \sum_{m=1}^{k} (1-\alpha r)^{k-m} \big( 1 + \alpha ( \lambda_i - r) \big)^{m-1}
	\label{eq:betasumk}
\end{eqnarray} 

Defining $\lambda_i = \zeta_i (1- \alpha r)$, we obtain 

\begin{equation}
(1-\alpha r)^{k-m} \big( 1 + \alpha ( \lambda_i - r) \big)^{m-1} = (1-\alpha r)^{k-1} (1 + \alpha \zeta_i)^{m-1} 
		\label{eq:xy} 
\end{equation} 

Writing eq.(\ref{eq:xy}) in eq.(\ref{eq:betasumk}) gives 

\begin{equation}  \label{eq:betazeta}
\beta_i(k) = \alpha \zeta_i (1-\alpha r)^{k} S(k) 
\end{equation} 

where $S(k)$ is 

\begin{equation}  \label{eq:Series}
S(k) = \sum_{m=1}^{k} (1 + \alpha \zeta_i)^{m-1} 
\end{equation}

Summing $-S(k)$ with $(1 + \alpha \zeta_i) S(k)$ yields 

\begin{equation}  \label{eq:S(k)}
S(k) = \frac{ (1 + \alpha \zeta_i)^{k} - 1 }{ \alpha \zeta_i }
\end{equation}

From eq.(\ref{eq:betazeta}), (\ref{eq:Series}) and (\ref{eq:S(k)}), we obtain  

\begin{equation}  \label{eq:betazeta1}
\beta_i(k) = (1-\alpha r)^{k} (1 + \alpha \zeta_i)^{k} - (1-\alpha r)^{k}
\end{equation}

Using the definition $\zeta_i = \lambda_i / (1- \alpha r)$ in  eq.(\ref{eq:betazeta1}) gives 

\begin{equation}  \label{eq:betazeta2} 
\beta_i(k) = \big( 1 + \alpha ( \lambda_i - r) \big)^{k} - (1-\alpha r)^{k} 
\end{equation} 

From eq.(\ref{eq:W_k}) and eq.(\ref{eq:betazeta2}), 

\begin{equation} \label{eq:W_kfromZeta}
{\bf M}^k = (1 - \alpha r)^k {\bf I} 
		+ \sum_{i=1}^{N} \big( 1 + \alpha ( \lambda_i - r) \big)^{k} {\bf V}_i 
			- \sum_{i=1}^{N} (1-\alpha r)^{k} {\bf V}_i 
\end{equation}

Let's put the $N$ eigenvalues of matrix ${\bf W}$ into two groups as follows: 
Let those eigenvalues which are smaller that $r$, 
belong to set $T = \{ \lambda_{j_t} \}_{j_t=1}^{N_t}$ where $N_t$ is the length of the set; and let 
those eigenvalues which are larger than $r$ belong to set $S=\{ \lambda_{j_s} \}_{j_s=1}^{N_s}$ 
where $N_s$ is the length of the set. We write the matrix ${\bf M}^k$ in eq.(\ref{eq:W_kfromZeta}) using 
this eigenvalue grouping 

\begin{equation} \label{eq:stm_M} 
{\mathbf M}^k  = {\bf M}_{tp}(k) + {\bf M}_{sp}(k) 
\end{equation} 

where 

\begin{equation} \label{eq:M1} 
{\bf M}_{tp}(k) = (1 - \alpha r)^k {\bf I} - \sum_{i=1}^{N} (1-\alpha r)^{k} {\bf V}_i 
		+ \sum_{j_t \in T} \big( 1 + \alpha ( \lambda_{j_t} - r) \big)^{k} {\bf V}_{j_t} 
\end{equation} 

and 

\begin{equation} \label{eq:M2} 
{\bf M}_{sp}(k) = \sum_{j_s \in S} \big( 1 + \alpha ( \lambda_{j_s} - r) \big)^{k} {\bf V}_{j_s} 
\end{equation}

We call the matrices ${\bf M}_{tp}(k)$ and ${\bf M}_{sp}(k)$ in (\ref{eq:M1}) and (\ref{eq:M2}) 
as transitory phase part and steady phase part, respectively, of the matrix ${\mathbf M}^k$. 

It's observed from eq.(\ref{eq:M1}) that the ${\bf M}_{tp}(k)$ converges to zero in a finite step number $k_T$ 
because relatively small step number $\alpha >0$ is chosen such that $(1-\alpha r) < 1$ and 
$1 + \alpha ( \lambda_{j_t} - r) < 1$. So, 

\begin{equation} \label{eq:nMtp} 
{\bf M}_{tp}(k) \approx {\bf 0},   \quad \quad k \geq k_T
\end{equation} 

Thus, what shapes the steady state behavior of the system in eq.(\ref{eq:solution}) and (\ref{eq:M}) is merely the 
${\bf M}_{sp}(k)$ in eq.(\ref{eq:M2} ). So, the steady phase solution is obtained from eq.(\ref{eq:solution}), 
(\ref{eq:M}) and (\ref{eq:M2}) using the above observations as follows 

\begin{eqnarray}  
{\bf x}_{sp}(k) & = & {\bf M}_{sp}(k)  {\bf x}(0) \label{eq:solutionMs} \\ 
	&  =  &	 \sum_{j_s \in S} \big( 1 + \alpha ( \lambda_{j_s} - r) \big)^{k} {\bf V}_{j_s} {\bf x}(0), 
	\quad \quad k \geq k_T \label{eq:solutionMs2}  
\end{eqnarray}

Let's define the interference vector, ${\bf J}_{sp}(k)$ as 

\begin{equation} \label{eq:J_intrf} 
{\bf J}_{sp}(k) = {\bf W} {\bf x}_{sp}(k) 
\end{equation} 

Using eq.(\ref{eq:symW}) in (\ref{eq:J_intrf}) and the orthonormal features in (\ref{eq:V_ortnorjMat}) yields 

\begin{equation} \label{eq:J_intrf_b} 
{\bf J}_{sp}(k) = 
\sum_{j_s \in S} \lambda_{j_s} \big( 1 + \alpha ( \lambda_{j_s} - r) \big)^{k} {\bf V}_{j_s} {\bf x}(0)
\end{equation}

First defining ${\bf V}_{j} {\bf x}(0) = {\bf u}_{j}$, 
and $\xi = \frac{\alpha}{1-\alpha r}$, 
then dividing vector ${\bf x}_{sp}(k)$ of eq.(\ref{eq:solutionMs2}) to ${\bf J}_{sp}(k)$ of 
eq.(\ref{eq:J_intrf_b}) elementwise and comparing the outcome with the "SIR" definition in 
eq.(\ref{eq:cirA}) results in 

\begin{eqnarray}  
\frac{x_{sp, i}(k)}{J_{sp, i}(k)} & = & \frac{1}{r} {\theta_i}(k),  
		\quad  \quad \quad \quad \quad i=1, \dots, N \label{eq:theta_xJ_i} \\ 
	 & = & 
	\frac{ \sum_{j_s \in S} ( 1 + \xi \lambda_{j_s} )^{k} u_{j_s, i} }{ \sum_{j_s \in S} \lambda_{j_s} ( 1 + \xi \lambda_{j_s} )^{k} u_{j_s, i} }
	\label{eq:theta_xJ_i2} 
\end{eqnarray}

In eq.(\ref{eq:theta_xJ_i2}), we assume that 
the ${\bf u}_{j} = {\bf V}_{j} {\bf x}(0)$ which corresponds to the eigenvector of the largest positive eigenvalue 
is different than zero vector. This means that we assume in the analysis here that 
${\bf x}(0)$ is not completely perpendicular to the mentioned eigenvector. 
This is something easy to check in advance. If it is the case, then 
this can easily be overcome 
by introducing a small random number to ${\bf x}(0)$ so that it's not completely perpendicular to the 
mentioned eigenvector.

From the analysis above, we observe that 

\begin{enumerate}

\item If all the (positive) eigenvalues greater than $r$ are the same, which is denoted as 
$\lambda_{b}$, then it's seen from (\ref{eq:theta_xJ_i2}) that 

\begin{eqnarray} \label{eq:theta_c} 
{\theta_{i}}(k) = \frac{r}{\lambda_{b}}, \quad i=1, \dots, N, \quad k \geq k_T
\end{eqnarray} 

\item  Similarly, if there is only one positive eigenvalue which is larger than $r$, 
shown as $\lambda_{b}$, then  eq.(\ref{eq:theta_c}) holds.

\item  If there are more than two different (positive) eigenvalues and the largest positive  
eigenvalue is single (not multiple), then  we see from (\ref{eq:theta_xJ_i}) that 
the term related to the largest (positive) eigenvalue dominates the sum of the nominator. 
Same observation is valid for the sum of the denominator. 
This is because a relatively small increase in $\lambda_{j}$ 
causes exponential increase as time step evolves, which is shown in the following: 
Let's show the two largest (positive) eigenvalues as $\lambda_{max}$ and $\lambda_{j}$ respectively and 
the difference between them as $\Delta \lambda$. So, $\lambda_{max} = \lambda_{j} + \Delta \lambda$. 
Let's define the following ratio between the terms related to the two highest eigenvalues 
in the nominator  

\begin{equation} \label{eq:expDeltaL}
K_n(k)  = \frac{ (1+\xi \lambda_{j})^{k} }{(1+\xi (\lambda_{j} + \Delta \lambda) )^{k}} 
\end{equation} 

where 

\begin{equation} \label{eq:xi} 
\xi = \frac{\alpha}{1-\alpha r} 
\end{equation} 

Similarly, let's define the ratio between the terms related to the two highest eigenvalues 
in the denominator as 

\begin{equation} \label{eq:expDeltaL_d}
K_d(k) = \frac{ \lambda_j (1+\xi \lambda_{j})^{k} }{ (\lambda_{j}+\Delta \lambda) (1+\xi (\lambda_{j} + \Delta \lambda) )^{k} }
\end{equation} 

From eq.(\ref{eq:expDeltaL}) and (\ref{eq:expDeltaL_d}), since $\frac{\lambda_{j}}{\lambda_{j} + \Delta} < 1$ 
due to the above assumptions,  

\begin{equation} \label{eq:K_dn} 
K_d(k) < K_n(k)
\end{equation}

We plot the ratio $K_n(k)$ in Fig. \ref{fig:ratioDeltaL} for some different $\Delta \lambda$ values and 
for a typical $\xi$ value.  
The Figure \ref{fig:ratioDeltaL} and eq.(\ref{eq:K_dn}) implies that 
the terms related to the $\lambda_{max}$ dominate 
the sum of the nominator and that of the denominator respectively. 
So, from eq.(\ref{eq:theta_xJ_i2}) and (\ref{eq:xi}), 

\begin{equation} \label{eq:theta_maxL} 
\frac{x_{sp, i}(k)}{J_{sp, i}(k)}  =  
\frac{ \sum_{j_s \in S} ( 1 + \xi \lambda_{j_s} )^{k} u_{j_s, i} }{ \sum_{j_s \in S} \lambda_{j_s} ( 1 + \xi \lambda_{j_s} )^{k} u_{j_s, i} }
\rightarrow 
\frac{ ( 1 + \xi \lambda_{max} )^{k} }{ \lambda_{max} ( 1 + \xi \lambda_{max} )^{k} } = \frac{1}{\lambda_{max}}, 
\quad k \geq k_T
\end{equation}

\item  If the largest positive eigenvalue is a multiple eigenvale, then, similarly, 
the corresponding terms in the sum 
of the nominator and that of the demoninator become dominant, which implies from eq.(\ref{eq:theta_xJ_i2}), 
(\ref{eq:expDeltaL}) and (\ref{eq:expDeltaL_d}) that 
$\frac{x_{sp, i}(k)}{J_{sp, i}(k)}$ converges to $\frac{1}{\lambda_{max}}$ as step number increases. 

\end{enumerate}

Using the observations 1 to 4, eq.(\ref{eq:nMtp}), the "SIR" definition 
in eq.(\ref{eq:cirA}), eq.(\ref{eq:theta_xJ_i}) and (\ref{eq:theta_xJ_i2}), 
we conclude that 

\begin{equation} \label{eq:cirA_sp1}
{\theta_i}(k) = \frac{ r x_{sp, i}(k) }{ \sum_{j = 1, j \neq i}^{N} w_{ij} x_{sp, j}(k) } = 
 	\frac{r}{\lambda_{max}},  	\quad k \geq k_T \quad  i=1, \dots, N,
\end{equation} 

where $\lambda_{max}$ is the largest (positive) eigenvalue of the matrix {\bf W}, and 
$k_T$ shows the finite time constant (during which the matrix ${\bf M}_{tp}(k)$ in eq.(\ref{eq:M1}) vanishes), 
which completes the proof.

\end{proof}

\vspace{0.2cm}

\emph{Definition:} Ultimate SIR value: In proposition 1, we showed that the SIR in (\ref{eq:cirA}) 
for every state in the autonomous discrete-time linear networks in eq.(\ref{eq:Diff_Linear}) converges to 
a constant value as step number goes to infinity.  We call this converged constant value 
as "ultimate SIR" and denote as ${\theta}^{ult}$. 

\vspace{0.2cm}

\subsection{ A more general autonomous linear discrete-time systems with symmetric matrix case
 \label{Subsection:MxGeneral} }

In this subsection, we analyse the following discrete-time autonomous linear system 

\begin{equation} \label{eq:generalDLS}
{\mathbf x}(k+1) = ( -\rho {\bf I} + {\bf W}  ) {\mathbf x}(k)
\end{equation}

where ${\bf I}$ is the identity matrix, $\rho$ is a positive real number, and 
$(-\rho {\bf I} + {\bf W})$ is the symmetric system matrix.  The real symmetric matrix ${\bf W}$ is shown 
in eq.(\ref{eq:matA_W_b}).

\vspace{0.2cm}
\emph{Proposition 2:}
\vspace{0.2cm}

In the the discrete-time linear system of eq.(\ref{eq:generalDLS}), let's assume that the spectral radius of 
symmetric matrix ${\bf W}$ in (\ref{eq:matA_W_b}) is larger than 1, i.e., the maximum of the norms of 
the eigenvalues is larger than 1. 


If $\rho$ is chosen such that 

\begin{enumerate}

\item
\begin{equation} 
0 < \rho  < 1,  \label{eq:generalDLS_condition} 
\end{equation} 

and 

\item

Define the eigenvalue(s) $\lambda_{m}$ as 

\begin{equation} \label{eq:lamb_m}
| \lambda_{m} - \rho | = \max  \{ | \lambda_{i} - \rho | \}_{i=1}^{N} > 1
\end{equation}

the eigenvalue $\lambda_{m}$ 
is unique. 
(In other words, $\rho$ is chosen in such a way that the equation eq.(\ref{eq:lamb_m}) does not hold 
for two eigenvalues with opposite signs. It would hold for a multiple eigenvalue as well, i.e., same sign.)

\end{enumerate}

then the defined "SIR" (${\theta_i}(k)$) in eq.(\ref{eq:cirA}) for any state $i$  
converges to the following ultimate SIR as step number $k$ evolves for any initial vector ${\bf x}(0)$ which is 
not completely perpendicular 
\footnote{
It's easy to check in advance if the initial vector ${\bf x}(0)$ is 
completely perpendicular to the eigenvector of the 
eigenvalue $\lambda_{m}$ in (\ref{eq:lamb_m}) of ${\bf W}$ or not. 
If this is the case, then this can easily be overcome 
by introducing a small random number to ${\bf x}(0)$ so that it's not completely perpendicular to the 
mentioned eigenvector. 
} 
to the eigenvector corresponding to the eigenvalue $\lambda_{m}$ in (\ref{eq:lamb_m}) of ${\bf W}$. 

\begin{equation} \label{eq:theta_const_general} 
\theta_i(k \geq k_T) =  \frac{\rho}{ \lambda_{m} }, \quad \quad i=1,2, \dots, N 
\end{equation} 

where $\lambda_{m}$ is the eigenvalue of the weight matrix ${\bf W}$ which satisfy 
eq.(\ref{eq:lamb_m}) and $k_T$ shows a finite time constant.

\begin{proof} 

From eq.(\ref{eq:generalDLS}), 

\begin{equation} \label{eq:genDLsolution} 
{\mathbf x}(k) = ( -\rho {\bf I} + {\bf W} )^{k} {\mathbf x}(0) 
\end{equation} 

where ${\bf x}(0)$ shows the initial state vector at step zero. 
Let's examine the powers of $( -\rho {\bf I} + {\bf W} )$ in (\ref{eq:genDLsolution}) in terms of 
the eigenvectors of ${\bf W}$ using eqs.(\ref{eq:symW})-(\ref{eq:V_ortnorjMat}):

\begin{equation} \label{eq:NW1} 
( -\rho {\bf I} + {\bf W} ) = - \rho {\bf I} + \sum_{i=1}^{N} \eta_i(1) {\bf V}_i 
\end{equation} 

where $\eta_i(1)$ is equal to 

\begin{equation} \label{eq:eta1} 
\eta_i(1) = \lambda_i 
\end{equation} 

The matrix $(-\rho {\bf I} + {\bf W})^2$ can be written as

\begin{equation} \label{eq:NW2}
(-\rho {\bf I} + {\bf W})^2 = \rho^2 {\bf I} + \sum_{i=1}^{N} \eta_i(2) {\bf V}_i
\end{equation}

where $\eta_i(2)$ is equal to

\begin{equation} \label{eq:eta2} 
\eta_i(2) = -\rho \lambda_i + (\lambda_i - \rho) \eta_i(1) 
\end{equation} 

Similarly, the matrix $(-\rho {\bf I} + {\bf W})^3$ can be written as 

\begin{equation} \label{eq:N3}
(-\rho {\bf I} + {\bf W})^3 = -\rho^3 {\bf I} + \sum_{i=1}^{N} \eta_i(3) {\bf V}_i
\end{equation}

where $\eta_i(3)$ is equal to

\begin{equation} \label{eq:eta3} 
\eta_i(3) = \rho^2 \lambda_i + (\lambda_i - \rho) \eta_i(2) 
\end{equation} 

So, at step $k$, the matrix $(-\rho {\bf I} + {\bf W})^k$ is obtained as 

\begin{equation} \label{eq:rhoW_k}
(-\rho {\bf I} + {\bf W})^k = (-\rho)^k {\bf I} + \sum_{i=1}^{N} \eta_i(k) {\bf V}_i
\end{equation}

where $\eta_i(k)$ is 

\begin{equation} \label{eq:eta_k} 
\eta_i(k) = (-\rho)^{k-1} \lambda_i + (\lambda_i - \rho) \eta_i(k-1) 
\end{equation} 

So, from (\ref{eq:eta1})-(\ref{eq:eta_k}) 

\begin{eqnarray}
\eta_i(1) &  = & \lambda_i   \label{eq:etasum1} \\
\eta_i(2) &  = & \lambda_i \Big( -\rho + ( \lambda_i - \rho) \Big)  \label{eq:etasum2} \\
\eta_i(3) &  = & 
\lambda_i \Big( \rho^2 - \rho ( \lambda_i - \rho) + ( \lambda_i - \rho)^2 \Big) \label{eq:etasum3} \\
           &  \vdots  &    \label{eq:etasumdots}    \\
\eta_i(k)   & = & \lambda_i \sum_{m=1}^{k} (-1)^{k-m} \rho^{k-m} ( \lambda_i - \rho)^{m-1}
        \label{eq:etasumk}
\end{eqnarray}

Defining $\lambda_i = \mu_i \rho$, we obtain

\begin{equation}  
\rho^{k-m} ( \lambda_i - \rho)^{m-1} = \rho^{k-1} (\mu_i -1)^{m-1}  \label{eq:xy2}
\end{equation}

Writing eq.(\ref{eq:xy2}) in eq.(\ref{eq:etasumk}) gives

\begin{equation}  \label{eq:betaeta}
\eta_i(k) = \lambda_i \rho^{k-1} S(k)
\end{equation}

where $S(k)$ is

\begin{equation}  \label{eq:Series2}
S(k) = \sum_{m=1}^{k} (-1)^{k-1} (\mu_i -1)^{m-1}
\end{equation}

Summing $S(k)$ with $(\mu_i -1) S(k)$ yields

\begin{equation}  \label{eq:S2}
S(k) = \frac{ (-1)^{k-1} + (\mu_i -1)^k }{ \mu_i }
\end{equation}

From eq.(\ref{eq:betaeta}), (\ref{eq:Series2}) and (\ref{eq:S2}), we obtain

\begin{equation}  \label{eq:betaeta1} 
\eta_i(k) = \lambda_i \rho^{k-1} \frac{ (-1)^{k-1} + (\mu_i -1)^k }{ \mu_i } 
\end{equation} 

Using the definition $\mu_i = \lambda_i / \rho$ in  eq.(\ref{eq:betaeta1}) gives

\begin{equation}  \label{eq:betaeta2}
\eta_i(k) = (-1)^{k-1} \rho^{k} + ( \lambda_i - \rho)^{k}
\end{equation}

From eq.(\ref{eq:rhoW_k}) and eq.(\ref{eq:betaeta2}),

\begin{equation} \label{eq:W_kfromEta}
(-\rho {\bf I} + {\bf W})^k = (-\rho)^k {\bf I} + \sum_{i=1}^{N} (-1)^{k-1} \rho^{k} {\bf V}_i 
               + ( \lambda_i - \rho)^{k} {\bf V}_i
\end{equation}

Let's put the $N$ eigenvalues of matrix ${\bf W}$ into two groups as follows:
Let those eigenvalues which satisfy $| \lambda_{j} - \rho | < 1$ 
belong to set $T = \{ \lambda_{j_t} \}_{j_t=1}^{N_t}$ where $N_t$ is the length of the set; and 
let all other eigenvalues (i.e. those which satisfy $| \lambda_{j} - \rho | > 1$) 
belong to set $S=\{ \lambda_{j_s} \}_{j_s=1}^{N_s}$
where $N_s$ is the length of the set. Here, $\rho$ is chosen such that 
no eigenvalue satisfy $| \lambda_{j} - \rho | = 1$. 
We write the matrix $(-\rho {\bf I} + {\bf W})^k$ in eq.(\ref{eq:W_kfromEta}) using
this eigenvalue grouping as follows 

\begin{equation} \label{eq:stm_N}
(-\rho {\bf I} + {\bf W})^k = {\bf N}_{tp}(k) + {\bf N}_{sp}(k)
\end{equation}

where

\begin{equation} \label{eq:N1}
{\bf N}_{tp}(k) = (-\rho)^k {\bf I} + \sum_{i=1}^{N} (-1)^{k-1} \rho^{k} {\bf V}_i 
             + \sum_{j_t \in T} ( \lambda_{j_t} - \rho)^{k} {\bf V}_{j_t} 
\end{equation}

and

\begin{equation} \label{eq:N2}
{\bf N}_{sp}(k) = \sum_{j_s \in S} ( \lambda_{j_s} - \rho)^{k} {\bf V}_{j_s} 
\end{equation}

In (\ref{eq:N1}), $| \lambda_{j_t} - \rho | < 1$ from the grouping mentioned above 
and $\rho$ is chosen such that $0 < \rho < 1$, which means that 
the ${\bf N}_{tp}(k)$ converges to zero in a finite step number $k_T$, i.e.,  

\begin{equation} \label{eq:nNtp}
{\bf N}_{tp}(k) \approx {\bf 0},   \quad \quad k \geq k_T
\end{equation}

Thus, what shapes the steady state behavior of the system in eq.(\ref{eq:genDLsolution})
is merely the ${\bf N}_{sp}(k)$ in eq.(\ref{eq:N2} ). 
We call the matrices ${\bf N}_{tp}(k)$ and ${\bf N}_{sp}(k)$ in (\ref{eq:N1}) and (\ref{eq:N2})
as transitory phase part and steady phase part, respectively, of the matrix ${\mathbf N}^k$.

So, the steady phase solution is obtained from eq.(\ref{eq:genDLsolution}), (\ref{eq:stm_N}) 
(\ref{eq:N1}) and (\ref{eq:N2}) as follows

\begin{eqnarray}
{\bf x}_{sp}(k) & = & {\bf N}_{sp}(k)  {\bf x}(0) \label{eq:solNs} \\
        &  =  &  
\sum_{j_s \in S} ( \lambda_{j_s} - \rho)^{k} {\bf V}_{j_s}{\bf x}(0), 
        \quad \quad k \geq k_T \label{eq:solNs2}
\end{eqnarray}

Let's define the interference vector, ${\bf J}_{sp}(k)$ as

\begin{equation} \label{eq:J_intrf2}
{\bf J}_{sp}(k) = {\bf W} {\bf x}_{sp}(k)
\end{equation}

Using eq.(\ref{eq:symW}) in (\ref{eq:J_intrf2}) and the orthonormal features in (\ref{eq:V_ortnorjMat}) yields

\begin{equation} \label{eq:J_intrf_b2}
{\bf J}_{sp}(k) = \sum_{j_s \in S} \lambda_{j_s} ( \lambda_{j_s} - \rho)^{k} {\bf V}_{j_s}{\bf x}(0) 
\end{equation}

Defining ${\bf V}_{j} {\bf x}(0) = {\bf u}_{j}$, and 
then dividing vector ${\bf x}_{sp}(k)$ of eq.(\ref{eq:solNs2}) to ${\bf J}_{sp}(k)$ of
eq.(\ref{eq:J_intrf_b2}) elementwise and comparing the outcome with the ``SIR'' definition in
eq.(\ref{eq:cirA}) results in 

\begin{eqnarray}
\frac{x_{sp, i}(k)}{J_{sp, i}(k)} & = & \frac{1}{r} {\theta_i}(k),
                \quad  \quad \quad \quad \quad i=1, \dots, N \label{eq:eta_xJ_i} \\
         & = & \frac{ \sum_{j_s \in S} ( \lambda_{j_s} - \rho)^{k} u_{j_s, i} }{ \sum_{j_s \in S} \lambda_{j_s} ( \lambda_{j_s} - \rho)^{k} u_{j_s, i} }      \label{eq:eta_xJ_i2}
\end{eqnarray}

In eq.(\ref{eq:eta_xJ_i2}), we assume that
the ${\bf u}_{j} = {\bf V}_{j} {\bf x}(0)$ corresponding to the eigenvector $\lambda_{m}$ in eq.(\ref{eq:lamb_m})
is different than zero vector. This means that we assume in the analysis here that
${\bf x}(0)$ is not completely perpendicular to the mentioned eigenvector.
This is something easy to check in advance. If it is the case, then
this can easily be overcome
by introducing a small random number to ${\bf x}(0)$ so that it's not completely perpendicular to the
mentioned eigenvector.


Here it's assumed that the eigenvalue $\lambda_{m}$ satisfying the equation eq.(\ref{eq:lamb_m}) is unique.
In other words, $\rho$ is chosen in such a way that the equation eq.(\ref{eq:lamb_m}) does not hold
for two eigenvalues with opposite signs. (It holds for a multiple eigenvalue, i.e., with same sign).
Using this assumption (which can easily be met by choosing $\rho$ accoridingly) 
in eq.(\ref{eq:eta_xJ_i2}) yields the following: 
The term related to eigenvalue $\lambda_{m}$ in eq.(\ref{eq:lamb_m}) 
dominates the sum of the nominator. 
This is because a relatively small decrease in $\lambda_{j}$ 
causes exponential decrease as time step evolves, which is shown in the following: 
Let's define the following ratio 

\begin{equation} \label{eq:expDeltaL2}
\kappa_n(k)  = \frac{ ( \lambda_{j_s} - \rho)^{k} }{ ( \lambda_{j_s} + \Delta \lambda - \rho)^{k} }
\end{equation} 

where $\Delta \lambda$ represents the decrease. Similarly, for denominator 

\begin{equation} \label{eq:expDeltaL_d2}
\kappa_d(k) = \frac{ \lambda_{j_s} ( \lambda_{j_s} - \rho)^{k} }{ (\lambda_{j_s} + \Delta \lambda)  ( \lambda_{j_s} + \Delta \lambda - \rho)^{k} }
\end{equation} 

From eq.(\ref{eq:expDeltaL2}) and (\ref{eq:expDeltaL_d2}), since $\frac{\lambda_{j}}{\lambda_{j} + \Delta} < 1$, 

\begin{equation} \label{eq:K_dn2} 
\kappa_d(k) < \kappa_n(k)
\end{equation}

\begin{figure}[htbp]
  \begin{center}
   \epsfxsize=30.0em    
\leavevmode\epsffile{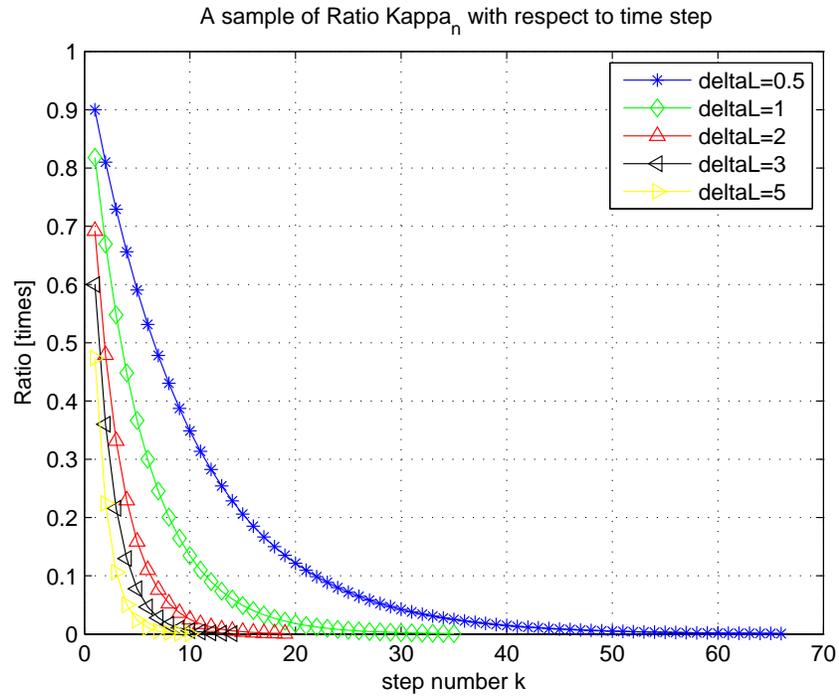}
   \vspace{-1em}        
  \end{center}
 \caption{ The figure shows same examples of ratio $\kappa_n$ in eq.(\ref{eq:expDeltaL2}) for some different
$\Delta \lambda$ values ($\lambda=5$). }
\label{fig:ratioDeltaL2}
\end{figure}

We plot some typical examples of the ratio $\kappa_n(k)$ in Fig. \ref{fig:ratioDeltaL2} for some 
different $\Delta \lambda$ values. 
The Figure \ref{fig:ratioDeltaL2} and eq.(\ref{eq:K_dn2}) implies that 
the terms related to the $\lambda_{m}$ dominate 
the sum of the nominator and that of the denominator respectively. 
So, from eq.(\ref{eq:eta_xJ_i2}) 

\begin{equation} \label{eq:eta_maxL} 
\frac{x_{sp, i}(k)}{J_{sp, i}(k)}  =  
\frac{ \sum_{j_s \in S} ( \lambda_{j_s} - \rho)^{k} u_{j_s, i} }{ \sum_{j_s \in S} \lambda_{j_s} ( \lambda_{j_s} 
- \rho)^{k} u_{j_s, i} } 
\rightarrow 
\frac{ ( \lambda_{m} - \rho)^{k} u_{j_m, i} }{ \lambda_{m} ( \lambda_{m} - \rho)^{k} u_{j_m, i} } = \frac{1}{\lambda_{max}}, 
\quad k \geq k_T
\end{equation} 

where $\lambda_{m}$ is defined by eq.(\ref{eq:lamb_m}).  
If the largest positive eigenvalue is a multiple eigenvale, then, similarly, 
the corresponding terms in the sum 
of the nominator and that of the demoninator become dominant, which implies from eq.(\ref{eq:eta_xJ_i2}), 
(\ref{eq:expDeltaL2})-(\ref{eq:eta_maxL}) that 
$\frac{x_{sp, i}(k)}{J_{sp, i}(k)}$ converges to $\frac{1}{\lambda_{max}}$ as step number evolves.

From eq.(\ref{eq:nNtp}), and the "SIR" definition 
in eq.(\ref{eq:cirA}), we conclude from eq.(\ref{eq:eta_xJ_i})-(\ref{eq:eta_maxL}) that 

\begin{equation} \label{eq:cirA_sp}
{\theta_i}(k) = \frac{ r x_{sp, i}(k) }{ \sum_{j = 1, j \neq i}^{N} w_{ij} x_{sp, j}(k) } = 
 	\frac{r}{\lambda_{m}},  	\quad k \geq k_T \quad  i=1, \dots, N,
\end{equation} 

where $\lambda_{m}$ is defined by eq.(\ref{eq:lamb_m}), and 
$k_T$ shows the finite time constant (during which the matrix ${\bf N}_{tp}(k)$ in eq.(\ref{eq:N1}) vanishes), 
which completes the proof.

\end{proof}


\section{Stabilized Discrete-Time Autonomous Linear Networks with Ultimate ``SIR''}
\label{Section:proposedNet} 

The proposed autonomous networks networks are 

\begin{enumerate}

\item

\begin{eqnarray} 
{\mathbf x}(k+1) & = & \Big( {\bf I} + \alpha ( -r {\bf I} + {\bf W} ) \Big) {\mathbf x}(k) 
                 \delta ( {\bf \theta}(k) - {\bf \theta}^{ult} )  \label{eq:SAL-USIR1} \\
	{\mathbf y}(k) & = & sign( {\mathbf x}(k) )    \label{eq:SAL-USIR1n}
\end{eqnarray}

where ${\bf W}$ is defined in (\ref{eq:matA_W_b}) 
$\alpha$ is step size, ${\bf I}$ is identity matrix and $r>0$ as in  eq.(\ref{eq:Diff_Linear}), 
${\bf \theta}(k) = [ \theta_1(k)  \theta_2(k)  \dots \theta_N(k) ]^T$, and
${\bf \theta}^{ult} = [ \theta_1^{ult} \theta_2^{ult}  \dots \theta_N^{ult} ]^T$,
and ${\mathbf y}(k)$ is the output of the network. In eq.(\ref{eq:SAL-USIR1}) 

\begin{equation} \label{eq:delta_fn}
{\bf \delta} ( {\bf \theta} - {\bf \theta}^{ult} ) = 
\left\{ 
\begin{array}{ll}
0 & \textrm{if and only if} \quad {\mathbf \theta}(k) = {\mathbf \theta}^{ult}, \\
1 & \textrm{otherwise} 
\end{array}
\right.
\end{equation} 

We call the network in ref.(\ref{eq:SAL-USIR1}) as Discrete 
Stabilized Autonomous Linear Networks by Ultimate ``SIR''1 (DSAL-U"SIR"1). 

\item 

\begin{eqnarray}
{\mathbf x}(k+1) & = &( -\rho {\bf I} + {\bf W}  ) {\mathbf x}(k) 
             \delta ( {\bf \theta}(k) - {\bf \theta}^{ult} )  \label{eq:SAL-USIR2} \\
          {\mathbf y}(k) & = & sign( {\mathbf x}(k) )    \label{eq:SAL-USIR2n}
\end{eqnarray}

where ${\bf I}$ is the identity matrix, $ 1 > \rho > 0$ and ${\bf W}$ is defined 
in eq.(\ref{eq:matA_W_b}), 
and ${\mathbf y}(k)$ is the output of the network. 
We call the network in ref.(\ref{eq:SAL-USIR2}) as DSAL-U"SIR"2.

\end{enumerate}

\vspace{0.2cm}
\emph{Proposition 3:} 
\vspace{0.2cm}

The proposed discrete-time networks of DSAL-U''SIR''1 in eq.(\ref{eq:SAL-USIR1}) 
is stable for any initial vector ${\bf x}(0)$ which is 
not completely perpendicular to the eigenvector corresponding to the largest eigenvalue of ${\bf W}$.
\footnote{ 
See the comments of proposition 1.  
}

\begin{proof} 
The proof of proposition 1 above shows that 
in the linear networks of eq.(\ref{eq:Diff_Linear}), 
the defined SIR in eq.(\ref{eq:cirA}) for state $i$  
converges to the constant ultimate SIR value in eq.(\ref{eq:theta_const}) 
for any initial condition ${\bf x}(0)$ within a finite step number $k_T$.
It's seen that the DSAL-U''SIR''1 in eq.(\ref{eq:Diff_Linear}) 
is nothing but the 
underlying network of the proposed networks SAL-U"SIR"1 without the $\delta$ function. 
Since the ``SIR'' in eq.(\ref{eq:cirA}) exponentially approaches to the constant Ultimate ``SIR'' 
in eq.(\ref{eq:theta_const}), the delta function eq.(\ref{eq:delta_fn}) will stop the exponential 
increase once ${\bf \theta}(k) = {\bf \theta}^{ult}$, at which the system outputs reach their steady 
state responses. So, the DSAL-U''SIR''1 is stable. 

\end{proof}

\vspace{0.2cm}
\emph{Proposition 4:} 
\vspace{0.2cm}

The proposed discrete-time network DSAL-U''SIR''2 in (\ref{eq:SAL-USIR2}) 
is stable for any initial vector ${\bf x}(0)$ which is 
not completely perpendicular to the eigenvector corresponding to the eigenvalue described in 
eq.(\ref{eq:lamb_m}). 
\footnote{ 
See the comments of proposition 2.  
}

\begin{proof} 
The proof of proposition 2 above shows that 
in the linear network of (\ref{eq:generalDLS}), 
the defined SIR in eq.(\ref{eq:cirA}) for state $i$  
converges to the constant ultimate SIR value in eq.
(\ref{eq:theta_const_general}),  
for any initial condition ${\bf x}(0)$ within a finite step number $k_T$.
It's seen that the linear networks of eq.(\ref{eq:generalDLS}) is nothing but the 
underlying network of the proposed network DSAL-U"SIR"2 without the $\delta$ function. 
Since the ``SIR'' in eq.(\ref{eq:cirA}) exponentially approaches to the constant Ultimate ``SIR'' 
in eq.(\ref{eq:theta_const_general}), the delta function eq.(\ref{eq:delta_fn}) will stop the exponential 
increase once ${\bf \theta}(k) = {\bf \theta}^{ult}$, at which the system output reach its steady 
state response. So, the DSAL-U''SIR''2 is stable. 

\end{proof}

\vspace{0.2cm}

So, from the analysis above for symmetric $\mathbf{ W }$ and $0 < r < \lambda_{max}$ for 
the SAL-U"SIR"1 in eq.(\ref{eq:SAL-USIR1}) and $0 < \rho < 1$ for the DSAL-U"SIR"2 in eq.(\ref{eq:SAL-USIR2}), 
 we conclude that  

\begin{enumerate}

\item 
The DSALU-''SIR''1 and DSALU-''SIR''2 does not show oscilatory behaviour because it's assured by the 
design parameter $r$ that $\rho$, respectively, that there is no eigenvalue on the unit circle.

\item 
The transition phase of the "unstable" linear network is shaped by the initial state vector 
and the phase space characteristics formed by the eigenvectors of ${\bf W}$. 
The network is stabilized by a switch function 
once the network has passed the transition phase. 
The output of the network then is formed taking the sign of the converged states. 
If the state converges to a plus or minus value is dictated by the phase space of the underlying linear network 
from the initial state vector at time 0.  
\end{enumerate}

Choosing the $r$ and $\rho$ in the DSALU-''SIR''1 and DSALU-''SIR''2 respectively such that the overall system 
matrix has positive eigenvalues makes the proposed networks exhibit 
similar features as Hopfield Network does as shown in the simulation results in section \ref{Section:SimuResults}.

As far as the design of weight matrix $r \mathbf{ I }$ ($\rho \mathbf{ I }$) and 
$\mathbf{ W }$ is concerned, well-known Hebb-learning rule (\cite{Hebb49}) is one of the commonly used methods 
(see e.g. \cite{Muezzinoglu04}). We proposed a method in \cite{Uykan08a} which is based 
on the Hebb-learning rule \cite{Hebb49}.  We summarize the design method here as well 
for the sake of completeness. 

\vspace{0.2cm}
\subsection{ Outer products based network design \label{Subsection:hebbianlearning} }
\vspace{0.2cm}

Let's assume that $L$ desired prototype vectors, $\{ \mathbf{ d }_s \}_{s=1}^{L}$, 
are given from  $(-1, +1)^N$. The proposed method is based on well-known Hebb-learning \cite{Hebb49} as follows: 

Step 1: Calculate the sum of outer products of the prototype vectors (Hebb Rule, \cite{Hebb49})

\begin{eqnarray} \label{eq:HebbQd}
\mathbf{ Q } = \sum_{s=1}^{L} \mathbf{ d }_s  \mathbf{ d }_s^T
\end{eqnarray}

Step 2: Determine the diagonal matrix $\bf{rI}$ and $\bf{W}$ as follows:

\begin{equation} \label{eq:rfromHebb}
r = q_{ii} + \vartheta
\end{equation}

where $\vartheta$ is a real number and   

\begin{equation} \label{eq:WfromHebb}
w_{ij} =
\left\{
\begin{array}{ll}
0 & \textrm{if} \quad i = j, \\
q_{ij} & \textrm{if} \quad i \neq j
\end{array}
\right.   \quad \quad i,j=1, \dots, N
\end{equation}

where $q_{ij}$ shows the entries of matrix $\mathbf{ Q }$, $N$ is the dimension of the vector 
$\mathbf{ x }$ and $L$ is the number of the prototype vectors ($N > L > 0$). 
From eq.(\ref{eq:HebbQd}), $q_{ii} = L$ since $\{ \mathbf{ d }_s \}$ is from $(-1, +1)^N$.

We assume that the desired 
prototype vectors are orthogonal and we use the following design procedure 
for matrices $\mathbf{ A }$, $\mathbf{ W }$ and $\mathbf{ b }$, which is based on 
Hebb learning (\cite{Hebb49}).

\vspace{0.2cm}
\emph{Proposition 5:} 
\vspace{0.2cm}

For the proposed network DSALU-"SIR"1 in eq.(\ref{eq:Diff_Linear}) whose weight matrix is designed by the 
proposed outer-products (Hebbian-learning)-based method above, if the prototype vectors are orthogonal, then 
the defined SIR in eq.(\ref{eq:cirA}) for any state converges to the 
following constant "ultimate SIR" awithin a finite step number 

\begin{equation} \label{eq:theta_Hebb_const}
\theta_i(k > k_T) =  \frac{r}{ N-L }
\end{equation}

where $N$ is the dimension of the network and $L$ is the prototype vectors, 
$t_T$ shows a finite step number for any initial condition ${\bf x}(0)$ which is not completely orthogonal 
to any of the raws of matrix $\mathbf{ Q }$ in eq.(\ref{eq:HebbQd}).  
\footnote{
It's easy to check in advance if the initial vector ${\bf x}(0)$ is 
completely  orthogonal 
to any of the raws of matrix $\mathbf{ Q }$ in eq.(\ref{eq:HebbQd}) or not.  
If so, then this can easily be overcome 
by introducing a small random number to ${\bf x}(0)$ so that it's not completely orthogonal to 
any of the raws of matrix $\mathbf{ Q }$. 
} 

\begin{proof} 

The proof is presented in Appendix I.

\end{proof}

\vspace{0.2cm}
\emph{Corollary 1:} 
\vspace{0.2cm}

For the proposed DSALU-"SIR"1 to be used as an associate memory system, whose 
weight matrix is designed by the 
proposed outer-products (Hebbian-learning)-based method in section \ref{Subsection:hebbianlearning} 
for $L$ orthogonal binary vectors of dimension $N$, 

\begin{equation} \label{eq:prop1and3}
\lambda_{max} = N-L 
\end{equation}

where $\lambda_{max}$ is the maximum (positive) eigenvalue of the weight matrix ${\bf W}$.


\vspace{0.2cm}

\begin{proof} 

From the proposition 1 and 4 above, the result in proposition 1 is valid for any real symmetric matrix 
${\bf W}$ whose maximum eigenvalue is positive, 
while the result of proposition 4 is for only the symmetric matrix designed 
by the method in section \ref{Subsection:hebbianlearning}. 
So, comparing the results of the proposition 1 and 4, we conclude that for the network in proposition 4,
the maximum (positive) eigenvalue of the weight matrix ${\bf W}$ is equal to $N-L$.

\end{proof}

\vspace{0.2cm}

\section{Simulation Results  \label{Section:SimuResults} }

We take similar examples as in \cite{Uykan08a}, \cite{Uykan08b} and \cite{Uykan09a} 
for the sake of brevity and easy reproduction of the simulation results.  
We apply the same Hebb-based (outer-products-based) design procedure (\cite{Hebb49}) in \cite{Uykan08a} and \cite{Uykan09a}, 
which is presented in section \ref{Subsection:hebbianlearning}. So, 
the weight matrix ${\bf W}$ in all the simulated networks (the proposed networks and Discrete-Time Hopfield Network) are 
the same. 

In this section, we present two examples, one with 8 neurons and one with 16 neurons. 
As in \cite{Uykan08b}, traditional Hopfield network is used a reference network.  
The discrete Hopfield Network \cite{Hopfield85} is 

\begin{equation} \label{eq:discreteHopfield}
{\bf x}^{k+1} = sign \Big( {\bf W} {\bf x}^{k}  \Big)
\end{equation} 

where ${\bf W}$ is the weight matrix and ${\bf x}^{k}$ is the state at time $k$, and at most one state is 
updated. 

\vspace{0.2cm}
\emph{Example 1:}
\vspace{0.2cm}
 
The desired prototype vectors are 

\begin{equation} \label{eq:ex1_D}
{\mathbf D} =
\left[
\begin{array}{c c c c c c c c}
1   &   1   & 1  &  1  & -1  &  -1  & -1  & -1  \\
1   &   1   & -1  &  -1  & 1  &  1  & -1  & -1  
\end{array}
\right]
\end{equation}

The weight matrices $r \bf{ I }$ and $\bf{ W }$, and the threshold vector $\bf{ b }$ 
are obtained as follows 
by using the outer-products-based design mentioned above 
and $\vartheta$ is chosen as -1 and for the DSALU-U''SIR''2 network, $\rho=0.5$.

\begin{equation} \label{eq:matA_W_b_ex1}
{\mathbf A} = 2{\mathbf I},  
\quad \quad
{\mathbf W} =
\left[
\begin{array}{c c c c c c c c}
0 & 2 & 0 & 0 & 0 & 0 &-2 &-2 \\
2 & 0 & 0 & 0 & 0 & 0 &-2 &-2 \\
0 & 0 & 0 & 2 &-2 &-2 & 0 & 0 \\
0 & 0 & 2 & 0 &-2 &-2 & 0 & 0 \\
0 & 0 &-2 &-2 & 0 & 2 & 0 & 0 \\
0 & 0 &-2 &-2 & 2 & 0 & 0 & 0 \\
-2 &-2 & 0 & 0 & 0 & 0 & 0 & 2 \\
-2 &-2 & 0 & 0 & 0 & 0 & 2 & 0
\end{array}
\right],  
\quad \quad
{\mathbf \nu} = {\mathbf 0}
\end{equation}

where ${\mathbf I}$ shows the identity matrix of dimension $N$ by $N$.

\begin{figure}[htbp]
  \begin{center}
   \epsfxsize=30.0em    
\leavevmode\epsffile{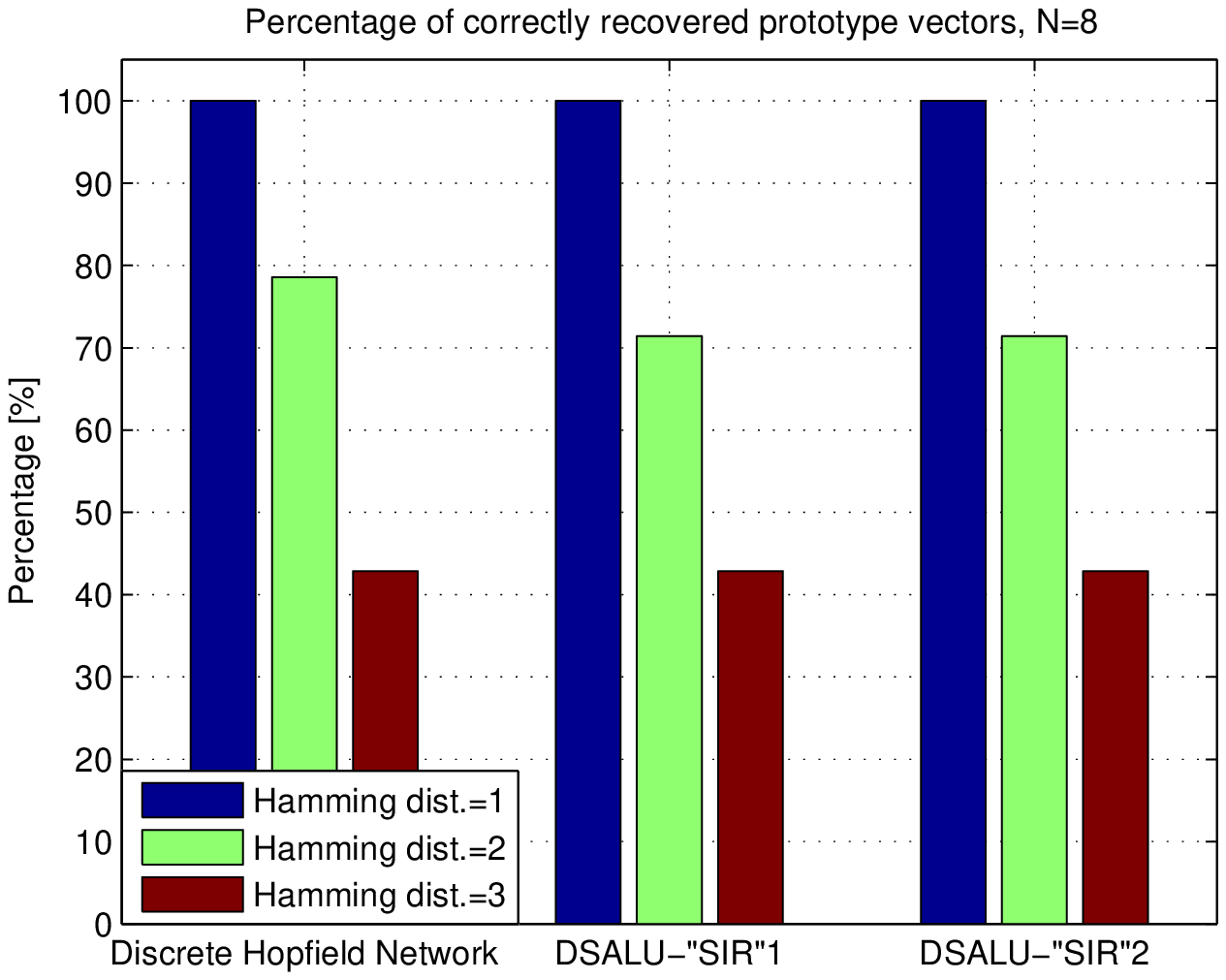}
   \vspace{-1em}        
  \end{center}
 \caption{ The figure
shows percentage of correctly recovered desired patterns for
all possible initial conditions in example 1 for the proposed DSALU-''SIR'' and Sign''SIR''NN
as compared to traditional Hopfield network with 8 neurons. }
\label{fig:DSALUSIR_ex1_percentage}
\end{figure}

The Figure \ref{fig:DSALUSIR_ex1_percentage} shows the percentages of correctly recovered desired patterns for 
all possible initial conditions $\mathbf{ x }(k=0) \in (-1,+1)^N$, in the proposed DSALU-"SIR"1 and 2 
as compared to traditional Hopfield network.  

Let $m_d$ show the number of prototype vectors and $C(N,K)$ represents the 
combination $N, K$ (such that $N \geq K \geq 0$), which is equal to $C(N,K)=\frac{N!}{(N-K)! K!}$, where $!$ shows factorial. 
In our simulation, the prototype vectors are from $(-1,1)^N$ as seen above. For initial conditions,  
we alter the sign of $K$ states where $K$=0, 1, 2, 3 and 4, which means the initial condition 
is within $K$-Hamming distance from the corresponding prototype vector. 
So, the total number of different possible combinations for the initial conditions for this example is 
24, 84 and 168 for 1, 2 and 3-Hamming distance cases respectively, which 
could be calculated by $m_d \times C(8,K)$, where $m_d =3$ and $K=$ 1, 2 and 3.

As seen from Figure \ref{fig:DSALUSIR_ex1_percentage}, the performance of the proposed networks DSALU-"SIR"1 and 2  
are the same as that of the discrete-time Hopfield Network for 1-Hamming distance case ($\%100$ for both networks) and 
are comparable results for 2 and 3-Hamming distance cases respectively.

\vspace{0.2cm}
\emph{Example 2:}
\vspace{0.2cm}

The desired prototype vectors are 

\begin{equation} \label{eq:ex2_D}
{\mathbf D} =
\left[
\begin{array}{c c c c c c c c c c c c c c c c}
1 & 1 & 1 & 1 & 1 & 1 & 1 & 1 & -1 & -1 & -1 & -1 & -1 & -1 & -1 & -1 \\
1 & 1 & 1 & 1 & -1 & -1 & -1 & -1 & 1 & 1 & 1 & 1 & -1 & -1 & -1 & -1 \\
1 & 1 & -1 & -1 & 1 & 1 & -1 & -1 & 1 & 1 & -1 & -1 & 1 & 1 & -1 & -1 
\end{array}
\right]
\end{equation}

The weight matrices $r{\bf I}$ and ${\bf W}$ and 
threshold vector ${\bf b}$ is obtained as follows 
by using the outer products based design as explained above. 
For matrix $r{\bf I}$, $\vartheta$ is chosen as -2.
The other network paramaters 
are chosen as in example 1. 

\begin{eqnarray} \label{eq:matA_W_b_ex2}
r {\mathbf I} & = & 3 {\mathbf I}, \nonumber \\
{\mathbf W} & = &
\left[
\begin{array}{c c c c c c c c c c c c c c c c}
0 &   3 &   1 &   1 &   1 &   1 &  -1 &  -1 &   1 &   1 &  -1 &  -1 &  -1 &  -1 &  -3 &  -3 \\
3 &   0 &   1 &   1 &   1 &   1 &  -1 &  -1 &   1 &   1 &  -1 &  -1 &  -1 &  -1 &  -3 &  -3 \\
1 &   1 &   0 &   3 &  -1 &  -1 &   1 &   1 &  -1 &  -1 &   1 &   1 &  -3 &  -3 &  -1 &  -1 \\
1 &   1 &   3 &   0 &  -1 &  -1 &   1 &   1 &  -1 &  -1 &   1 &   1 &  -3 &  -3 &  -1 &  -1 \\
1 &   1 &  -1 &  -1 &   0 &   3 &   1 &   1 &  -1 &  -1 &  -3 &  -3 &   1 &   1 &  -1 &  -1 \\
1 &   1 &  -1 &  -1 &   3 &   0 &   1 &   1 &  -1 &  -1 &  -3 &  -3 &   1 &   1 &  -1 &  -1 \\
-1 &  -1 &   1 &   1 &   1 &   1 &   0 &   3 &  -3 &  -3 &  -1 &  -1 &  -1 &  -1 &   1 &   1 \\
-1 &  -1 &   1 &   1 &   1 &   1 &   3 &   0 &  -3 &  -3 &  -1 &  -1 &  -1 &  -1 &   1 &   1 \\
1 &   1 &  -1 &  -1 &  -1 &  -1 &  -3 &  -3 &   0 &   3 &   1 &   1 &   1 &   1 &  -1 &  -1 \\
1 &   1 &  -1 &  -1 &  -1 &  -1 &  -3 &  -3 &   3 &   0 &   1 &   1 &   1 &   1 &  -1 &  -1 \\
-1 &  -1 &   1 &   1 &  -3 &  -3 &  -1 &  -1 &   1 &   1 &   0 &   3 &  -1 &  -1 &   1 &   1 \\
-1 &  -1 &   1 &   1 &  -3 &  -3 &  -1 &  -1 &   1 &   1 &   3 &   0 &  -1 &  -1 &   1 &   1 \\
-1 &  -1 &  -3 &  -3 &   1 &   1 &  -1 &  -1 &   1 &   1 &  -1 &  -1 &   0 &   3 &   1 &   1 \\
-1 &  -1 &  -3 &  -3 &   1 &   1 &  -1 &  -1 &   1 &   1 &  -1 &  -1 &   3 &   0 &   1 &   1 \\
-3 &  -3 &  -1 &  -1 &  -1 &  -1 &   1 &   1 &  -1 &  -1 &   1 &   1 &   1 &   1 &   0 &   3 \\
-3 &  -3 &  -1 &  -1 &  -1 &  -1 &   1 &   1 &  -1 &  -1 &   1 &   1 &   1 &   1 &   3 &   0 
\end{array}
\right],   \nonumber \\
{\mathbf \nu} & = & {\mathbf 0} 
\end{eqnarray}

\begin{figure}[htbp]
  \begin{center}
   \epsfxsize=30.0em    
\leavevmode\epsffile{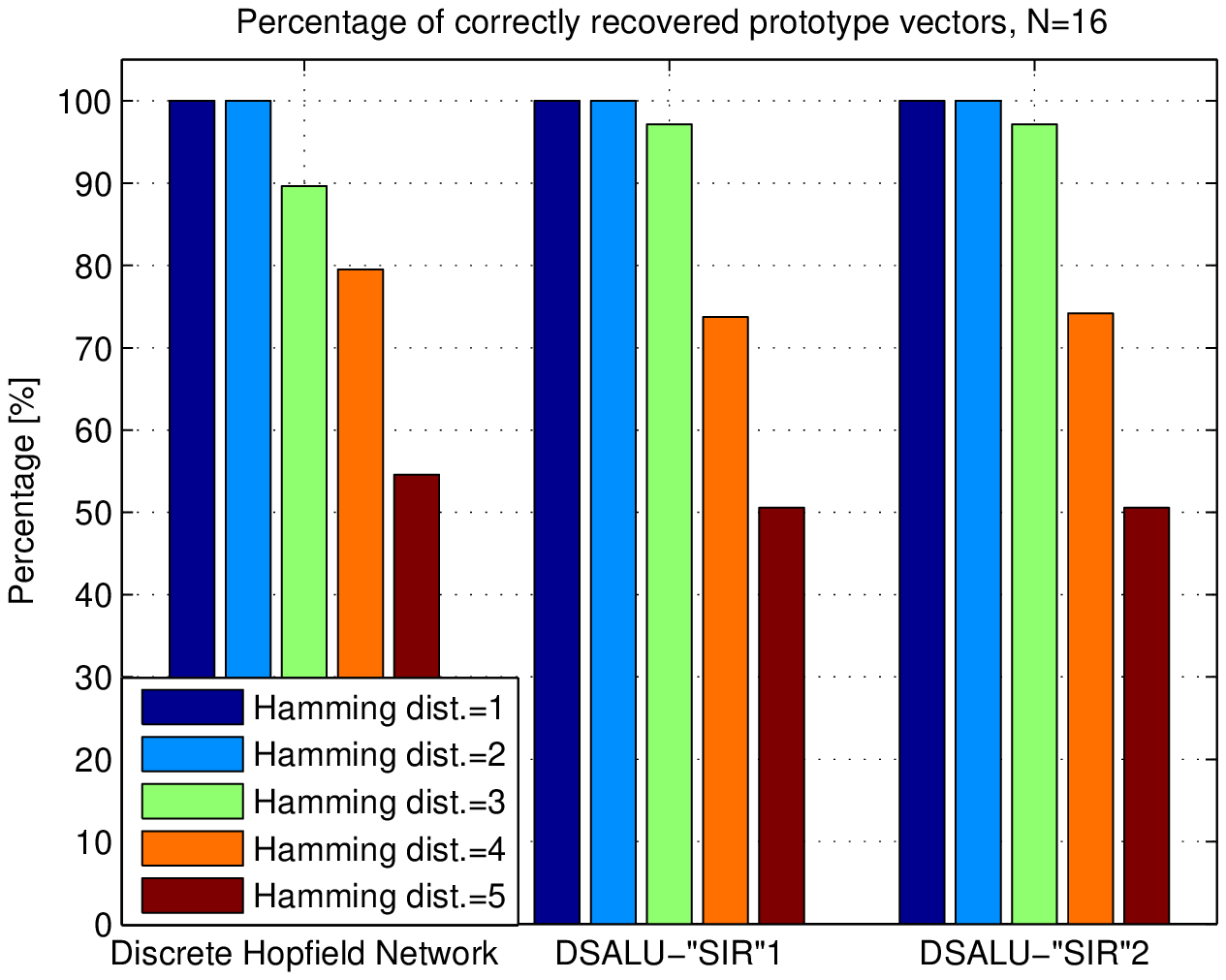}
   \vspace{-1em}        
  \end{center}
 \caption{ The figure 
shows percentage of correctly recovered desired patterns for 
all possible initial conditions in example 2 for the proposed DSALU-"SIR" and Sign"SIR"NN 
as compared to traditional Hopfield network with 16 neurons. }
\label{fig:DSALUSIR_ex2_percentage}
\end{figure}

The Figure \ref{fig:DSALUSIR_ex2_percentage} shows the percentages of correctly recovered desired patterns for 
all possible initial conditions $\mathbf{ x }(k=0) \in (-1,+1)^{16}$, in the proposed DSALU"SIR"1 and 2 
as compared to the traditional Hopfield network.  

The total number of different possible combinations for the initial conditions for this example is
64, 480 and 2240 and 7280 for 1, 2, 3 and 4-Hamming distance cases respectively, which
could be calculated by $m_d \times C(16,K)$, where $m_d =4$ and $K=$ 1, 2, 3 and 4. 

As seen from Figure \ref{fig:DSALUSIR_ex2_percentage} the performance of the proposed networks DSALU-"SIR"1 and 2  
are the same as that of Hopfield Network for 1 and 2-Hamming distance cases ($\%100$ for both networks), and 
are comparable for 3,4 and 5-Hamming distance cases respectively.

\section{Conclusions  \label{Section:CONCLUSIONS}}

Using the same ``SIR'' concept as in \cite{Uykan08a}, and \cite{Uykan08b},
we, in this paper, analyse the ``SIR'' of the states in the following two
$N$-dimensional discrete-time autonomous linear systems:

\begin{enumerate}

\item  The system ${\mathbf x}(k+1) = \big( {\bf I} + \alpha ( -r {\bf I} + {\bf W} ) \big) {\mathbf x}(k)$
which is obtained by discretizing the autonomous continuous-time
linear system in \cite{Uykan09a} using Euler method; where
${\bf I}$ is the identity matrix, $r$ is a positive real number, and
$\alpha >0$ is the step size.

\item A more general autonomous linear system descibed by ${\mathbf x}(k+1) = -\rho {\mathbf I + W} {\mathbf x}(k)$,
where ${\mathbf W}$ is any real symmetric matrix whose diagonal elements are zero, and
${\bf I}$ denotes the identity matrix and $\rho$ is a positive real number.

\end{enumerate}

Our analysis shows that:

\begin{enumerate}

\item The ``SIR'' of any state
converges to a constant value, called ``Ultimate SIR'', in a finite time
in the above-mentioned discrete-time linear systems.

\item The ``Ultimate SIR'' in the first system above is equal to $\frac{\rho}{\lambda_{max}}$ where
$\lambda_{max}$ is the maximum (positive) eigenvalue of the matrix ${\bf W}$.
These results are in line with those of \cite{Uykan09a} where corresponding continuous-time linear system is examined.

\item The ``Ultimate SIR'' in the second system above is equal to $\frac{\rho}{\lambda_{m}}$ where
$\lambda_{m}$ is the eigenvalue of ${\bf W}$ which satisfy
$| \lambda_{m} - \rho | = \max  \{ | \lambda_{i} - \rho | \}_{i=1}^{N}$ if
$\rho$ is accordingly determined from the interval $0 < \rho < 1$
as described in (\ref{eq:lamb_m}).

\end{enumerate}

In the later part of the paper, we use the introduced ``Ultimate SIR'' to stabilize the (originally unstable) networks.
It's shown that the proposed Discrete-Time
``Stabilized''-Autonomous-Linear-Networks-with-Ultimate-SIR'' exhibit features which are
generally attributed to Discrete-Time Hopfield Networks.
Taking the sign of the converged states, the proposed networks are applied to binary associative memory design.
Computer simulations show the effectiveness of the
proposed networks as compared to traditional discrete Hopfield Networks.

As far as design of the design of the weight matrices are concerned, we also present
an outer-products (Hebbian-learning)-based method, and show that if the prototype vectors are orthogonal
in the proposed DSAL-U''SIR''1 network,
then the ultimate SIR $\theta^{ult}$ is equal to $\frac{r}{ N-L }$
where $N$ is the dimension of the network and $L$ is the prototype vectors.

\vspace{0.2cm}

%

\section*{Appendix I}

\vspace{0.2cm}
\emph{Proof of Proposition 5:} 
\vspace{0.2cm}


The solution of the proposed network DSALU-"SIR"1 in eq.(\ref{eq:Diff_Linear}) is

\begin{equation} \label{eq:solution_a}
{\mathbf x}(k) = \Big( {\bf I} + \alpha ( -r {\bf I} + {\bf W} ) \Big)^{k} {\mathbf x}(0) 
\end{equation}

Let's denote the system matrix as 

\begin{eqnarray} \label{eq:MIW}
\mathbf{ M } = {\bf I} + \alpha ( -r {\bf I} + {\bf W} )
\end{eqnarray}

From eq.(\ref{eq:HebbQd}) and (\ref{eq:WfromHebb}), 

\begin{eqnarray} \label{eq:WQL}
\mathbf{ W } = \mathbf{ Q } - L \mathbf{ I }
\end{eqnarray}

where $L$ is the number of orthogonal prototype vector. Using eq.(\ref{eq:WQL}) in (\ref{eq:MIW}) gives 

\begin{eqnarray} \label{eq:rW1}
\mathbf{ M } = \Big( 1 - \alpha (r+L) \Big) \mathbf{ I } + \alpha \mathbf{ Q } 
\end{eqnarray} 

and since $\mathbf{ d }_s \in (-1,+1)^N$, 

\begin{eqnarray} \label{eq:Q2} 
\mathbf{ Q }^2 = N \mathbf{ Q }
\end{eqnarray}

where $N$ is the dimension of the system, i.e., the number of states, and $\mathbf{Q}$ is given in (\ref{eq:HebbQd}).  

Next, we examine the powers of matrix ${\bf M})$ since the solution of the system is 
${\bf x}(k) = \mathbf{ M }^{k} {\bf x}(0)$:

First let's define $b$ and $c$ as follows 

\begin{eqnarray} 
b =  1 - \alpha (r+L)  \label{eq:rz} \\
c =  1 - \alpha (r+L-N)  \label{eq:rz2}
\end{eqnarray}

From eq.(\ref{eq:rW1}) and (\ref{eq:rz}), 

\begin{eqnarray} \label{eq:rW1b}
\mathbf{ M } = b \mathbf{ I } + \sigma(1) \mathbf{ Q } 
\end{eqnarray} 

where 

\begin{equation} \label{eq:gamma1}
\sigma(1) = \alpha 
\end{equation}

The matrix ${\bf M}^2$ is 

\begin{equation} \label{eq:rW2}
{\bf M}^2 = b^2 {\bf I} + \sigma(2) {\bf Q}
\end{equation}

where $\sigma(2)$ is equal to 

\begin{equation} \label{eq:gamma2} 
\sigma(2) = \alpha ( b + c ) 
\end{equation} 

Similarly, the matrix ${\bf M}^2$ is obtained as 

\begin{equation} \label{eq:rW3}
{\bf M}^3 = b^3 {\bf I} + \sigma(3) {\bf Q}
\end{equation}

where $\sigma(3)$ is 

\begin{equation} \label{eq:gamma3}
\sigma(3) = \alpha ( b^2 + bc + c^2) 
\end{equation}

For $k=4$, 

\begin{equation} \label{eq:rW4}
{\bf M}^4 = b^4 {\bf I} + \sigma(4) {\bf Q}
\end{equation}

where $\sigma(4)$ is 

\begin{equation} \label{eq:gamma4}
\sigma(4) = \alpha ( b^3 + b^2 c + bc^2 + c^3 ) 
\end{equation}

So, when we continue, we observe that the $k$'th power of the matrix ${\bf M}$ is obtained as 

\begin{equation} \label{eq:rWk}
{\bf M}^k = b^k {\bf I} + \sigma(k) {\bf Q} 
\end{equation}

where $\sigma(k)$ is 

\begin{equation} \label{eq:gamma_k}
\sigma(k) = \alpha \sum_{m=1}^{k} b^{k-m} c^{m-1} 
\end{equation} 

where $b = 1 - \alpha (r+L)$ and $c = 1 - \alpha (r+L-N)$ as defined in eq.(\ref{eq:rz}) and (\ref{eq:rz2}), 
respectively.  Let's define the following constant $\varphi$ 

\begin{equation}  \label{eq:mu_bc}
\varphi = \frac{b}{c} = \frac{1 - \alpha (r+L)}{1 - \alpha (r+L-N)}
\end{equation}

Using (\ref{eq:mu_bc}) in (\ref{eq:gamma_k}) results in 

\begin{equation} \label{eq:gamma_k_mu}
\sigma(k) = \alpha c^{k-1} \sum_{m=1}^{k} \varphi^{k-m}
\end{equation} 

Summing $-\sigma(k)$ with $\varphi \sigma(k)$ yields 

\begin{equation}  \label{eq:gamma_mu}
\sigma(k) = \frac{\varphi^k -1}{\varphi-1} \alpha c^{k-1}
\end{equation} 

From eq.(\ref{eq:rWk}) and (\ref{eq:gamma_mu}), the matrix ${\bf M}^k$ is written as follows  

\begin{equation} \label{eq:rMbc} 
{\bf M}^k = b^k {\bf I} + \alpha c^{k-1} \frac{\varphi^k -1}{\varphi-1} {\bf Q}
\end{equation} 

Using the definition of $b$ and $c$ in eq.(\ref{eq:rz}) and (\ref{eq:rz2}), respectively, in  (\ref{eq:rMbc}) gives 

\begin{equation} \label{eq:rM12}
{\bf M}^k = {\bf M}_{tp}(k) + {\bf M}_{sp}(k)
\end{equation}

where 

\begin{equation} \label{eq:nM1} 
{\bf M}_{tp}(k) = \Big( 1 - \alpha (r+L) \Big)^k {\bf I} - \frac{1}{\alpha N} \Big( 1 - \alpha (r+L) \Big)^k {\bf Q}
\end{equation} 

and 

\begin{equation} \label{eq:nM2} 
{\bf M}_{sp}(k) = \frac{ \Big( 1 - \alpha (r+L-N) \Big)^k }{\alpha N} {\bf Q}
\end{equation}

In above equations, the number of network dimension ($N$) is much larger than the number of prototype vector ($L$), i.e. 
$N >> L$. In Hopfield networks, theoretically, $L$ is in the range of \%15 of $N$ (e.g. \cite{Muezzinoglu05}).  
So, $N > r+L$  by choosing $r$ accordingly. The learning factor positive $\alpha$ is also typically a 
relatively small number less than 1. Therefore, $\Big( 1 - \alpha (r+L) \Big) < 1$ and 
$\Big( 1 - \alpha (r+L-N) \Big)>1$.  This means that 1) the ${\bf M}_{tp}(k)$ in eq.(\ref{eq:nM1}) 
vanishes (aproaches to zero) within a finite step number $k_T$; and 
2) what shapes the steady state behavior of the system is merely ${\bf M}_{sp}$.

\begin{equation} \label{eq:nMtp2} 
{\bf M}_{tp}(k) \approx {\bf 0},   \quad \quad k \geq k_T
\end{equation} 

We call the matrices ${\bf M}_{tp}(k)$ and ${\bf M}_{sp}(k)$ in (\ref{eq:nM1}) and (\ref{eq:nM2}) 
as transitory phase part and steady phase part, respectively, of the matrix ${\mathbf M}^k$. 

So, the steady phase solution is obtained from eq.(\ref{eq:solution}), 
(\ref{eq:rM12}) and (\ref{eq:nM2}) 

\begin{eqnarray}  
{\bf x}_{sp}(k) & = & {\bf M}_{sp}(k)  {\bf x}(0) \label{eq:nsolMs} \\ 
	&  =  &	 \frac{ \Big( 1 - \alpha (r+L-N) \Big)^k }{\alpha N} {\bf Q} {\bf x}(0), 
	\quad \quad k \geq k_T \label{eq:nsolMs2}  
\end{eqnarray}

Let's define the interference vector, ${\bf J}_{sp}(k)$ as 

\begin{equation} \label{eq:J_intrf_n} 
{\bf J}_{sp}(k) = {\bf W} {\bf x}_{sp}(k) 
\end{equation} 

From eq.(\ref{eq:WQL}), (\ref{eq:Q2}) and (\ref{eq:J_intrf_n}) 

\begin{eqnarray} 
{\bf J}_{sp}(k) & = & (\mathbf{ Q } - L \mathbf{ I }) {\bf x}_{sp}(k)    \label{eq:J_intf_n} \\
                & = & \frac{ \Big( 1 - \alpha (r+L-N) \Big)^k }{\alpha N} (N-L) {\bf Q} {\bf x}(0)  \label{eq:J_intf_n2} 
\end{eqnarray} 

So, dividing vector ${\bf x}_{sp}(k)$ of eq.(\ref{eq:nsolMs2}) to ${\bf J}_{sp}(k)$ of 
eq.(\ref{eq:J_intf_n}) elementwise and comparing the outcome with the "SIR" definition in 
eq.(\ref{eq:cirA}) results in

\begin{equation} \label{eq:ntheta_i} 
{\theta_i}(k) = \frac{r}{N-L},  	\quad \quad i=1, \dots, N
\end{equation}

which completes the proof.


\vspace{0.2cm}

\section*{Acknowledgments}

This work was supported in part by Academy of Finland and Research Foundation (Tukis\"{a}\"{a}ti\"{o}) of Helsinki 
University of Technology, Finland.

\nocite{*}
\bibliographystyle{IEEE}

%

\vspace{3cm}



\newpage

\listoffigures

\newpage

\begin{figure}[htbp]
  \begin{center}
   \epsfxsize=30.0em    
\leavevmode\epsffile{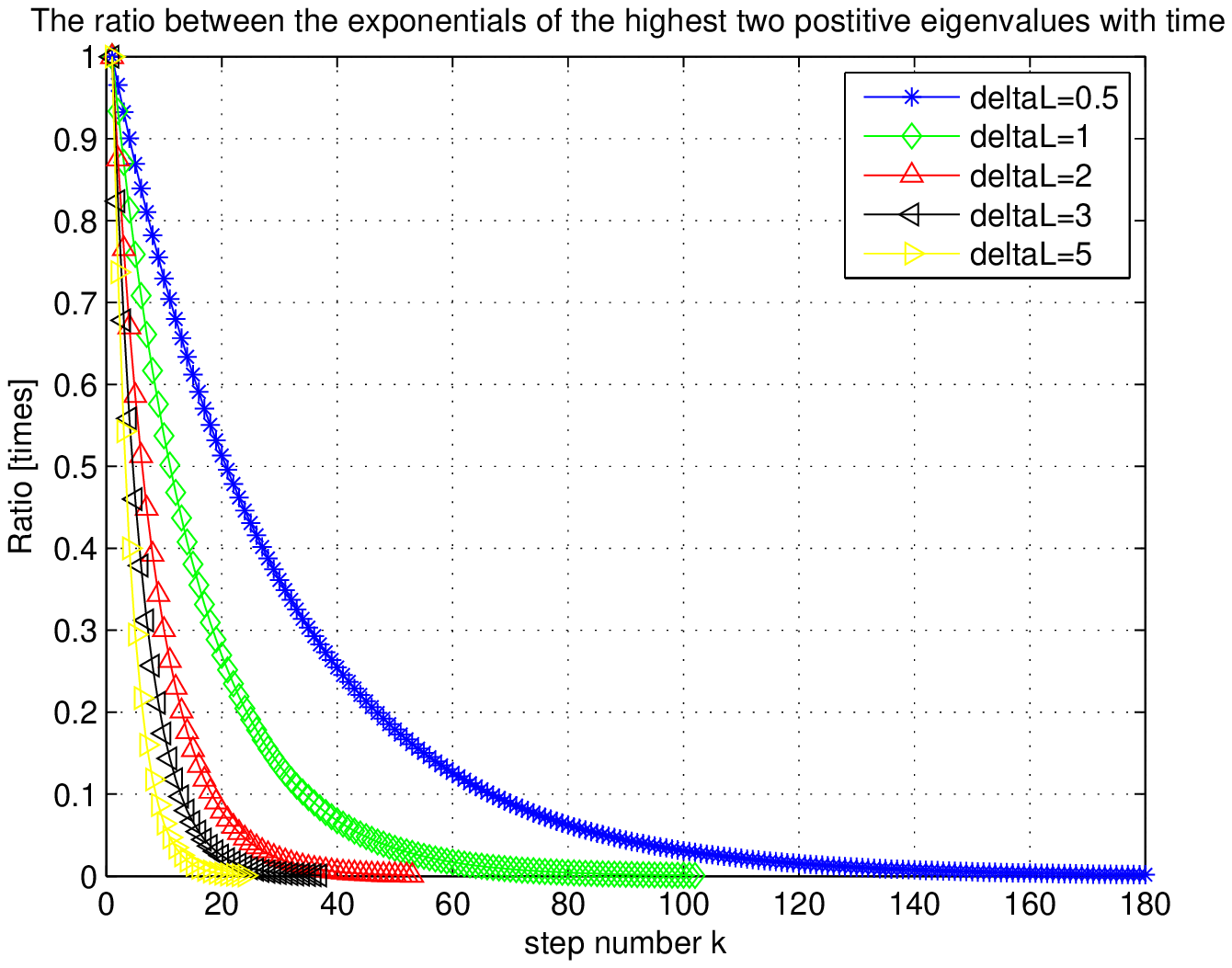}
   \vspace{-1em}        
  \end{center}
 \caption{ The figure shows the ratio $K_n$ in eq.(\ref{eq:expDeltaL}) for some different 
$\Delta \lambda$ values ($\lambda=5, \xi=0.11$). }
\label{fig:ratioDeltaL}
\end{figure}

\newpage

\newpage

\newpage

\end{document}